\documentclass[10pt, a4paper]{article}

\usepackage[]{lrec2026} 
\usepackage{mathtools}
\usepackage{booktabs}
\usepackage{amssymb}
\usepackage{multirow}
\usepackage{orcidlink}

\usepackage{listings}
\title{Demystifying Funding: Reconstructing a Unified Dataset of the UK Funding Lifecycle}

\name{William Thorne~\textsuperscript{$1,2$}~\orcidlink{0000-0002-8947-6261}, Rupert Shepherd~\textsuperscript{$1$}~\orcidlink{0000-0002-8283-904X}, Diana Maynard~\textsuperscript{$2$}~\orcidlink{0000-0002-1773-7020}} 

\address{\textsuperscript{$1$}~National Gallery, \textsuperscript{$2$}~University of Sheffield \\
         \{wthorne1,d.maynard\}@sheffield.ac.uk\\
         rupert.shepherd@nationalgallery.org.uk}

\abstract{
We present a reconstruction of UKRI's Gateway to Research (GtR) database that links funding opportunities to their resulting project proposals through panel meeting outcomes. Unlike existing work that focuses primarily on funded projects and their outcomes, we close the complete funding lifecycle by integrating three previously disconnected data sources: the GtR project database, UKRI funding opportunities, and competitive funding decision records across UKRI's research councils. We describe the technical challenges of data collection, including navigating inconsistent publication formats and restricted access to panel decisions. The resulting dataset enables a holistic interrogation of the entire funding process, from opportunity announcement to research outcomes. We release the database and associated code. \\ \newline \Keywords{scientific funding, Gateway to Research, information extraction, data integration}}

\begin{document}

\maketitleabstract

\section{Introduction}
\label{section:introduction}

The UK Research and Innovation (UKRI) Gateway to Research (GtR) database provides the most complete public record of UK-funded research projects, encompassing project metadata, participant information, institutional affiliations, and research outcomes. While GtR has enabled valuable meta-research, it suffers from well-documented quality issues. \citet{carlin_tracking_2025} find that a quarter of reported software outputs have no links and 45\% have missing or erroneous URLs. \citet{liyanage_fairness_2024} further note frequent duplicate data where "awards cross multiple years and/or were awarded to more than one institution." UKRI itself acknowledges "limitations on the accuracy, reliability and completeness of the data" \citep{ukri_gtr_guide_2026}, noting that classifications are inconsistently applied across funders, organisations lack unique identifiers across constituent systems, and historic information is not consistently comprehensive. A recent National Audit Office report found that data limitations are restricting UKRI's ability to manage its investments strategically \citep{nao_ukri_2025}.

Beyond data quality, GtR suffers from issues of coverage as it maintains no public relationship between funded projects and the funding opportunities that solicited them. This omission is of particular concern as opportunities mark UKRI's first point of involvement in many projects' lifecycles and define the assessment criteria against which proposals are evaluated \citep{thorne2026evaluatingllmbasedgrantproposal}. Compounding this, panel meeting outcomes that determine which proposals receive funding are published inconsistently across research councils, employing a mixture of PDFs and spreadsheets with variable structure and detail. Most concerning, two major councils publish their data exclusively through a third-party dashboard with data exports disabled, despite ostensibly operating under CC BY-NC-SA 4.0 licensing. This lack of transparency is notable given that bias in peer review and funding decisions has been well documented \citep{nesta_bias_2018}, and that research into funding process fairness overwhelmingly focuses on the final stages precisely because the earlier stages remain opaque \citep{edicaucus_2024}.

We argue that by limiting accessibility to funding decision data, UKRI actively discourages scrutiny of awards at the point where final funding decisions are made \citep{liyanage_fairness_2024}. This work addresses this gap by capturing the full funding lifecycle, from opportunity publication through panel decision to research outcomes, integrating three previously disconnected data sources into a unified database that extends existing GtR data.\footnote{Our code and data can be found here: \url{https://github.com/wrmthorne/GtR-Extended}}

\section{Related Work}
\label{section:related-work}

\subsection{GtR as a Research Data Source}

GtR serves as a primary data source for a range of meta-research applications. \citet{vanino_knowledge_2019} used GtR to assess business performance effects of publicly-funded R\&D grants, finding that firms participating in UKRI-funded projects experienced higher employment and turnover growth. GtR+ is a recent initiative that links GtR data with external sources such as OpenAlex and Companies House to support the analysis of topics including the impact of rail connectivity on research collaboration and the regional distribution of research funding \citep{irc_gtr_2025}. However, both projects inherit GtR's lifecycle gap. Our work is complementary, providing the upstream link between opportunities, panel decisions, and awards to enable analysis across the full funding pipeline.

\subsection{Data Quality and Reporting Concerns}

\citet{carlin_tracking_2025} found that artifact sharing in GtR remains low, highlighting the dependence of GtR outcome data on the Researchfish platform, which until recently served as UKRI's primary source for outcomes data. UKRI's own documentation acknowledges that outcome reporting was not mandatory prior to December 2009, that classifications are inconsistently applied across research councils, and that the same organisation may appear under multiple variant names \citep{ukri_gtr_guide_2026}. Project regions are based on lead applicant postcodes, potentially misrepresenting where research occurs.

\subsection{Funding Fairness and Transparency}

\citet{liyanage_fairness_2024} identified systematic over-representation of Russell Group institutions in UKRI and significant under-funding of Post-92 universities, despite the latter serving larger and more socially diverse student populations; however, their analysis was necessarily limited to successful awards as GtR records only funded projects. Our dataset extends to include the most complete public information on unfunded, pending, and deferred proposals. The EDI-Caucus found that 44\% of the literature focuses on gender bias, with racial inequity and institutional prestige receiving comparatively little attention, and that "little is known about the biases that affect which proposals reach review panels in the first place" \citep{edicaucus_2024}. \citet{nesta_bias_2018} similarly document how confirmation bias and tribalism in peer review can disadvantage novel or unconventional proposals. By providing structured access to panel-level decisions along side opportunities and project outcomes, our dataset enables the kind of cross-stage analysis that these studies identify as lacking.

\subsection{Linking Funding Opportunities, Meetings and Outcomes}

To our knowledge, no prior work has linked UKRI funding opportunities to their resulting proposals and panel decisions across research councils. Our work fills this gap, enabling research questions that span the full funding pipeline.

\section{Data Sources}
\label{section:data-sources-and-selection}

Our dataset integrates three primary sources: the UKRI Gateway to Research database, the complete collection of UKRI funding opportunities, and competitive funding decision records for each research council.

\begin{figure*}
    \centering
    \includegraphics[width=\linewidth]{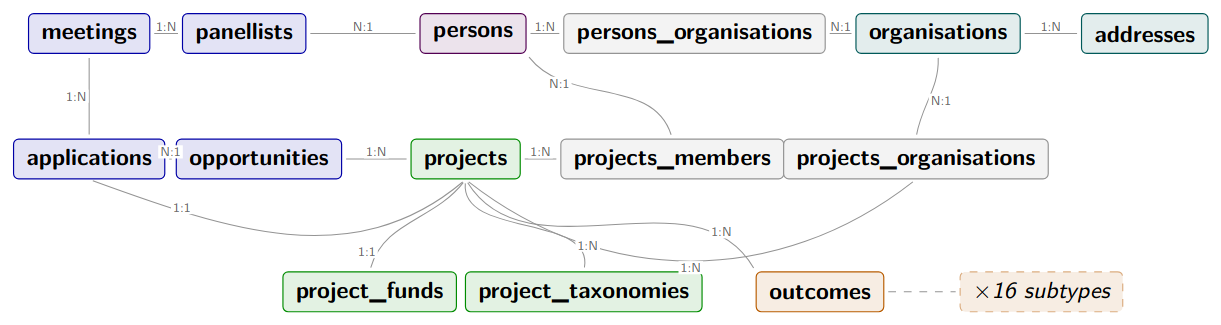}
    \caption{Simplified database schema. Tables are coloured according to primary entity: meeting/panel/opportunity (blue), project (green), people (purple), organisations (teal), outcomes (orange), link tables (grey).}
    \label{fig:database-schema}
\end{figure*}

\subsection{Gateway to Research}
\label{section:data-sources-gtr}

We extract data via all available GtR endpoints, including projects, funding records, persons, organisations, and outcome types (publications, collaborations, disseminations, policy influences, intellectual property, spin-outs, among others), ingesting into a PostgreSQL database schema centred on the \textit{Project} entity. Outcomes are implemented using joined table inheritance with a generic base table and outcome-specific child tables. Given schema mismatches with GtR's data structures, ingestion proceeds with constraints disabled; we report dangling references in \S~\ref{section:data-extraction}. Where source IDs were available we reused them; otherwise we construct deterministic UUID5 identifiers.

\subsection{Funding Opportunities}
\label{section:data-sources-opportunities}

We retrieved all pending, open, and closed opportunities from the UKRI opportunity finder, constructing a structured table of core metadata: title, funders, publication/opening/closing dates, award values, and status. As status and closing date change over time, this represents a snapshot at the time of writing. Not all funded projects correspond to a published opportunity; research councils also operate responsive mode schemes and formula-based allocation mechanisms \citep{ukri_crcrm_2024, ukri_gtr_guide_2026}.

Raw HTML was parsed into a tree-structured representation: the opportunity summary forms the root, with accordion sections (e.g. ``What we are looking for'', ``How to apply'') as children, each storing an ordered sequence of paragraphs, lists, and subheadings as further subtrees. Updates are extracted into date-text pairs. This structure is serialised to a consistent markdown representation for downstream use with language models, and serves as the basis for the hierarchical retrieval described in \S~\ref{section:opportunity-metadata-extraction}.

\subsection{Panel Meeting Outcomes and Attendance}
\label{section:data-sources-meetings}

\begin{table}
\centering
    \begin{tabular}{lrrrr}
    \toprule
    Council & Meets. & Apps. & Panels & Recon. \\
    \midrule
    AHRC  &   449 &  9,112 & 2,102 & 283 \\
    BBSRC &    31 &  5,215 & 3,109 &  20 \\                                         
    EPSRC &   308 &  4,557 & 2,554 & 228 \\
    ESRC  &   158 &  4,992 &   395 &  10 \\
    MRC   &   324 &  7,522 &   339 &  16 \\
    NERC  &   153 &  7,966 & 3,231 &  32 \\
    STFC  &    27 &  1,145 &     0 &   0 \\
    \midrule
    \textbf{Total} & \textbf{1,450} & \textbf{40,509} & \textbf{11,730} & \textbf{589} \\
    \bottomrule
    \end{tabular}
    \caption{Statistics of panel meeting and attendance extraction for each council. The table records the number of meetings, applications, successful outcomes, individual panel appearances (times a person was recorded present at a meeting), and meetings with successfully reconciled panel attendance.}
    \label{tab:meeting-stats}                                             
\end{table}

Collecting, and extracting outcomes and attendance from each of the councils was a significant undertaking given the inconsistent publication format and location, both by council and over time. Statistics for extracted data can be found in Table~\ref{tab:meeting-stats}. ESRC is the most accessible, publishing to a continuously updated spreadsheet on the UKRI website; however, panel attendance is not available for all meetings. Most other councils' data was sourced from a combination of the UKRI site and the Government Web Archive. Consistency and file format varies wildly; for example, pre-2025 MRC consistently relies on visual encoding in xslx files, using cell background colour to indicate funding status and column header numbers for scores, while NERC predominantly uses tables in PDFs that are devoid of any standardised layout or content. In most instances, attendance is recorded separately from application outcomes, if at all, and with insufficient information to reliably reconcile in many cases. Most problematically, EPSRC, post-2018 AHRC\footnote{Pre-2018 outcomes are available via the National Archives.}\footnote{Unavailable at the time of writing following our report to UKRI concerning unintentionally published applicant names, affiliations and project titles for unsuccessful proposals.}, and MRC from 2025 onward publish exclusively through Tableau, all with data exports disabled.

We argue this accessibility barrier goes against the spirit of UKRI's CC BY-NC-SA 4.0\footnote{\url{https://www.ukri.org/who-we-are/terms-of-use/}} licensing and is particularly problematic from the perspective of interrogating the funding process, given that these meetings mark the final determination of which projects receive funding.

\section{Metadata Extraction from Opportunities}
\label{section:opportunity-metadata-extraction}

Much information about funding opportunities exists only in unstructured textual content. While some opportunities list award ranges in a summary table, the UKRI contribution percentage, project duration, and total fund value are frequently only stated within prose, distributed across document sections. We frame this as a closed-domain question-answering task: for each metadata field, retrieve the most relevant passages and extract a structured answer. Documents have a mean of $2,824$ whitespace-delimited tokens ($4,317$ excluding Innovate UK who typically publish short summaries of their full documents from the Innovation Funding Service), with the longest spanning $15,126$ words, motivating experimentation with retrieval-based context reduction \citep{paulsen_context_2025}.

\begin{table}
\centering
\small
\begin{tabular}{llcccc}
\toprule
\multicolumn{2}{c}{Configuration} & & \multicolumn{2}{c}{Overall} \\
\cmidrule(lr){1-2} \cmidrule(lr){4-5}
Chunker & $\alpha$ & RR & Acc (\%) & Unk (\%) \\
\midrule
    Baseline & --- &  & 81.8 & 2.6 \\
    \addlinespace
    Simple (best) & 1.00 & \checkmark & 82.9 & 4.5 \\
    \addlinespace
    Hierarchical & 0.00 & \checkmark & 84.0 & 2.2 \\
    Hierarchical & 0.25 &            & 84.4 & 3.7 \\
    Hierarchical & \textbf{0.25} & \checkmark & \textbf{87.0} & \textbf{2.6} \\
    Hierarchical & 0.50 & \checkmark & 83.6 & 2.6 \\
    Hierarchical & 1.00 &            & 84.0 & 4.1 \\
\bottomrule
\end{tabular}
\caption{Extraction accuracy for selected retrieval configurations. Baseline passes the full opportunity text. RR = re-ranker. The best simple chunker configuration and all hierarchical configurations exceeding the baseline are shown; full results are shown in Appendix~\ref{appendix:full-results}.}
\label{tab:extraction-results-condensed}
\end{table}

\subsection{Hierarchical Retrieval}
\label{section:hierarchical-extraction}

We exploit the tree-structured document representation as a natural chunking mechanism. Each node yields one chunk comprising its header and all descendant content, such that a level-2 chunk subsumes all level-3 content within it. Chunks are scored against a query using a three-stage hybrid pipeline: a weighted combination of BM25 \citep{robertsonProbabilisticRelevanceFramework2009} and dense cosine similarity (via \href{https://huggingface.co/sentence-transformers/all-MiniLM-L6-v2}{all-MiniLM-L6-v2} \citep{wang2020minilm}, normalised to $[0,1]$):

\begin{equation}
S_{\text{hybrid}}(q, c) = \alpha \cdot S_{\text{BM25}}(q, c) + (1 - \alpha) \cdot S_{\theta}(q, c)
\end{equation}

\noindent The top-$k$ ($k=10$) candidates are re-ranked by a cross-encoder (\href{https://huggingface.co/cross-encoder/ms-marco-MiniLM-L6-v2}{ms-marco-MiniLM-L-6-v2} \citep{wang2020minilm}). We then apply \textit{branch-deduplication}: for any ancestor--descendant pair among the candidates, the lower-scoring chunk is discarded. The top-$k_{\text{final}}$ ($k_{\text{final}}=5$) surviving chunks are concatenated as retrieval context.

\subsection{Evaluation}
\label{section:extraction-evaluation}

We evaluate against manually annotated metadata from 101 opportunities, stratified first by funding council (Figure~\ref{fig:eval-council-distribution}) and then by document length to ensure coverage across UKRI's heterogeneous publication styles. Up to six fields are annotated for each opportunity: minimum and maximum award value, total funding, UKRI contribution percentage, and minimum and maximum funding duration. This resulted in 489 question-answer pairs. We compare our hierarchical chunker against a sliding-window baseline (200-word chunks, 50-word overlap), vary $\alpha \in \{0, 0.25, 0.5, 0.75, 1.0\}$ to assess the contribution of each retrieval component, and ablate the re-ranker from each configuration.

Table~\ref{tab:extraction-results-condensed} summarises extraction accuracy for selected configurations (full per-field results in Appendix~\ref{appendix:full-results}). The full-document baseline achieves 85.3\% accuracy, demonstrating that the extraction task is tractable even without retrieval. However, the baseline benefits from seeing all relevant context simultaneously; its main disadvantage is cost and latency at scale.

The simple (sliding-window) chunker underperforms the baseline in all configurations, with accuracy dropping as low as 76.9\% when the re-ranker is disabled at $\alpha{=}1$ (pure BM25). Without the re-ranker, increasing $\alpha$ degrades accuracy monotonically (83.4\% at $\alpha{=}0$ to 76.9\% at $\alpha{=}1$), suggesting that BM25 alone produces poor chunk rankings for these documents, likely because keyword overlap between the question and irrelevant sections (e.g.\ eligibility criteria mentioning funding amounts in passing) misleads lexical retrieval. Re-ranking consistently recovers performance for simple chunking (e.g.\ from 83.4\% to 82.8\% at $\alpha{=}0$, and from 76.9\% to 82.2\% at $\alpha{=}1$), confirming that the initial retrieval set contains relevant chunks but ranks them poorly.

Hierarchical chunking outperforms both alternatives, with 8 of 10 configurations exceeding the baseline. We attribute this to chunking at section boundaries, which preserves the natural co-occurrence of related information. The best configuration ($\alpha{=}0.25$, re-ranker disabled) achieves 87.7\% overall accuracy, a 2.4 percentage point improvement over the full-document baseline. Notably, the re-ranker has a mixed effect under hierarchical chunking: it improves accuracy at $\alpha{=}0$ (83.6\% $\to$ 87.1\%) but slightly harms it at $\alpha{=}0.25$ (87.7\% $\to$ 86.9\%), suggesting that the hybrid scoring already produces a strong ranking when the dense and BM25 signals are appropriately balanced. Maximum award remains the hardest field across all configurations (best: 66.7\%), which we attribute to frequent confusion with the closely related total funding field (see \S\ref{section:error-analysis}).

\subsection{Error Analysis}
\label{section:error-analysis}

To better understand model failure modes, we classify all 60 errors produced by the best-performing configuration (hierarchical, $\alpha{=}0.25$, without re-ranker; 87.7\% accuracy). Errors fall into three categories: \textit{false positives} (47), where the model returns a value when the ground truth is null; \textit{value mismatches} (9), where the model retrieved value does not align with the human annotation; and \textit{false negatives} (4), where the model returns null despite an answer being present in the text.

False positives dominate, accounting for 78\% of all errors. We further subdivide these into \textit{field confusion} (17/47) and \textit{hallucination} (30/47). Field confusion occurs when the model returns the correct value for a different field, for example, returning the total funding amount when asked for the maximum award, or the maximum duration when asked for the minimum. The most common confusion pairs are maximum$\to$minimum duration (9 cases) and total funding$\to$maximum award (4 cases). In the case of total funding, much of the confusion arises from the ambiguous phrasing of ``[candidate] can apply for a share of up to [total funding]'' in Innovate UK opportunities. While semantically allowable for a single applicant to be awarded the full total funding amount, this does not happen in reality.

The remaining 30 false positives are hallucinations in which the model produces a plausible-looking value that does not appear in the source document. Of these, 26 are entirely fabricated, 3 can be traced to values present in the document's metadata table rather than the body text, and 1 appears elsewhere in the document body in an unrelated context.

Value mismatches (9 cases) typically involve rounding differences in monetary amounts or minor duration disagreements. False negatives are rare (4 cases), most often arising when the relevant information is spread across multiple non-adjacent paragraphs, suggesting that the retrieval context occasionally fails to capture all relevant passages.

\section{Data Extraction and Alignment}
\label{section:data-extraction}

\begin{table}
    \centering
    \small
    \begin{tabular}{lrr}
        \toprule
        Entity & Count \\
        \midrule
        Projects & 173,220 \\
        People & 141,833 \\
        Organisations & 89,884 \\
        Outcomes & 2,619,446 \\
        \addlinespace
        Meetings & 1,449 \\
        Panel Appearances & 6,951 \\
        Applications & 38,862 \\
        Opportunities & 2,053 \\
        \bottomrule
    \end{tabular}
    \caption{Number of captured entities, grouped by source.}
    \label{tab:entity-counts}
\end{table}

Table~\ref{tab:entity-counts} summarises the entities captured. All projects have titles, abstracts, and funding records, and 2.6 million outcomes are linked to projects with no orphaned references. We find 1,520 orphaned project member records and 10,128 orphaned project partner records, consistent with UKRI's acknowledgement that organisations lack unique identifiers across constituent systems \citep{ukri_gtr_guide_2026}. A further 22,519 projects have no associated team members, likely reflecting incomplete historical data.

\subsection{Entity Linking}
\label{section:entity-linking}

Because the three data sources share no common identifiers, we perform a series of linking steps to connect them into a unified dataset. All thresholds were chosen conservatively to favour precision over recall.

\paragraph{Application to Project.} Applications are linked to GtR projects by exact case-insensitive match of the application's \texttt{application\_id} or \texttt{award\_id} against a project's \texttt{grant\_reference}. This linked 10,531 of 38,862 applications (27.1\%). Cross-validation on linked pairs with available metadata shows 97.2\% organisation agreement (N=2,930), 95.7\% PI surname agreement (N=3,074), and median title similarity of 1.000 (mean 0.996, N=3,048), yielding an estimated precision of 98.0\% (95\% CI: 97.7-98.3\%). The remaining 72.9\% of unlinked applications primarily reflect unfunded proposals with no corresponding GtR record.

\paragraph{Application to Opportunity.} Applications are linked to opportunities via fuzzy title matching with multiple constraints. Candidate text is the \texttt{opportunity\_name} from the application, falling back to the meeting name. Fuzzy ratio scores are computed against the opportunity title. Candidates are filtered by council-funder agreement and date ordering (earliest meeting on or after opportunity opening date). The score is boosted by 0.1 when the application route matches the opportunity funding type, and penalised by 0.15 when the total awarded amount falls outside the opportunity's award range. A minimum score threshold of 0.65 is required. This linked 11,514 applications (29.6\%) to an opportunity.

\paragraph{Panel Attendance to Person.} Panel attendance records are linked to GtR persons in two phases. First, records are disambiguated by clustering within each council on (surname, initial), splitting sub-clusters when full first names conflict or organisation similarity falls below 0.7. Second, each cluster is aligned to GtR persons with project membership records: unique surname-initial matches are linked directly; where multiple candidates exist, the best organisation similarity must exceed 0.6 with a margin of 0.15 over the second best. This linked 4,474 of 6,951 panel attendance records (64.4\%) to a GtR person, with 100\% surname containment and 71.3\% organisation agreement on linked records.

\section{Conclusion}
\label{section:conclusions}

We have presented a reconstruction of UKRI's Gateway to Research database that closes the funding lifecycle by linking opportunities, proposals, and panel decisions, enabling analysis at the point where final funding decisions are made and supporting future work on allocation fairness \citep{liyanage_fairness_2024} and research policy evaluation. Our work also documents concerning barriers to access, particularly the restrictions on panel meeting data \citep{nao_ukri_2025} -- data that enables investigation of panel composition and potential bias. By incorporating unfunded applications we provide a broader context around opportunity competitiveness and the full project lifecycle. We release our database and accompanying code to support further scrutiny of UK public research funding.

\section{Ethical Considerations}
\label{section:ethics}

All data used in this work is publicly available under CC BY-NC-SA 4.0. The dataset contains no personal data beyond what is already publicly disclosed by UKRI, and individuals are identified only through their institutional roles as principal investigators or panellists.

\section{Limitations}
\label{section:limitations}

Our dataset represents a snapshot at the time of collection; opportunity statuses and project details may have since updated. Meeting and panel alignment is incomplete as alignment was performed conservatively to avoid introducing incorrect data. Application-project linking covers 27.1\% of applications, with the remainder primarily reflecting unfunded proposals with no GtR record. Application-opportunity linking covers 29.6\% via fuzzy title matching with heuristic thresholds that may introduce errors. Finally, GtR's own data quality limitations, including orphaned records and inconsistent historical reporting, are inherited by our dataset.

\section{Acknowledgements}
\label{section:acknowledgements}

This work was supported by the Arts and Humanities Research Council [grant number AH/X004201/1].

\section{Bibliographical References}\label{sec:reference}

\bibliographystyle{lrec2026-natbib}
\bibliography{lrec2026-example}

\appendix

\section{Meetings by Council Over Time}

\begin{figure*}
    \centering
    \includegraphics[width=\linewidth]{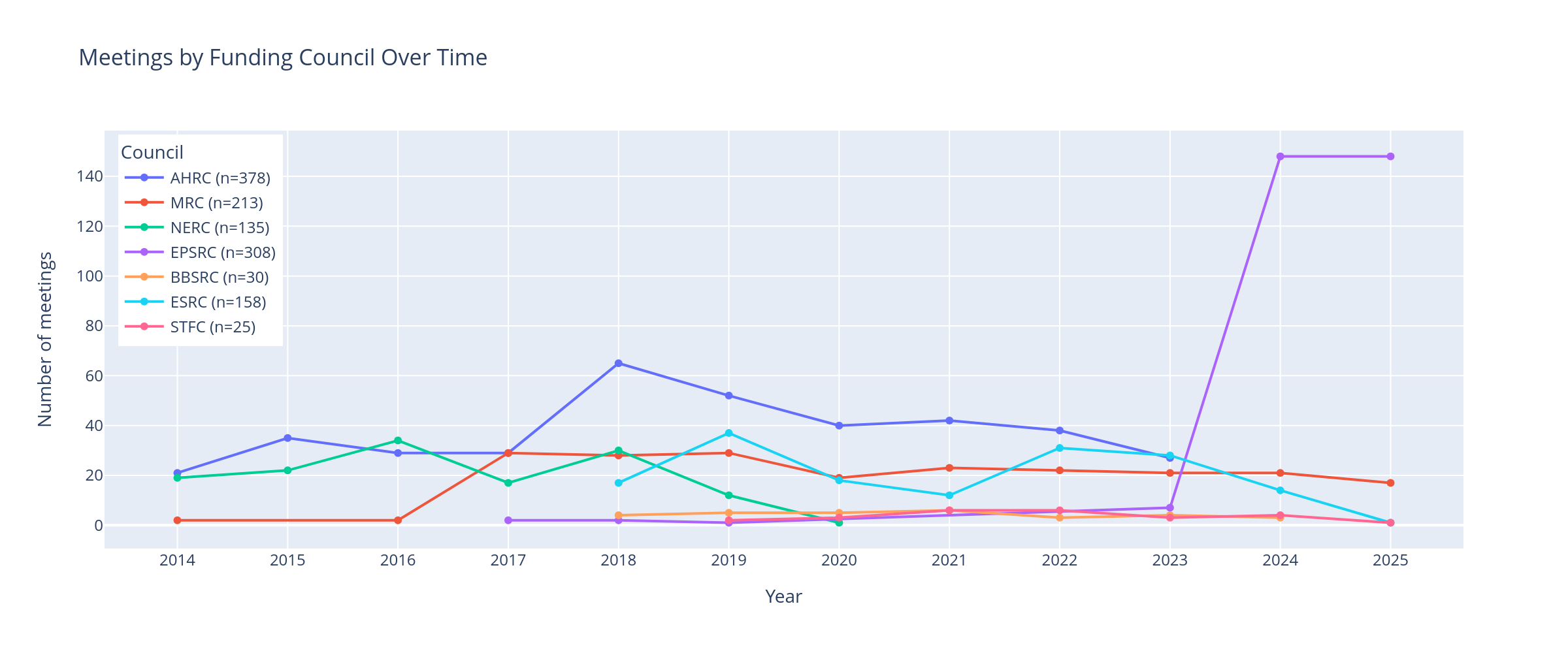}
    \caption{Number of meetings extracted per council over time.}
    \label{fig:meetings-per-council}
\end{figure*}

\section{Metadata Extraction Prompt}

\begin{table*}
    \centering
    \begin{tabular}{c|c}
        \textbf{Metadata Field} & \textbf{Question} \\
        \hline
        \textit{minimum\_award} & What is the minimum fund value (£)? \\
        \textit{maximum\_award} & What is the maximum fund value (£)? \\
        \textit{total\_funding} & What is the total fund value (£) split among successful applications? \\
        \textit{funding\_percentage} & What percentage (\%) of the project's funding will UKRI fund? \\
        \textit{minimum\_funding\_duration} & What is the minimum duration of the project/funding? \\
        \textit{maximum\_funding\_duration} & What is the maximum duration of the project/funding? \\
    \end{tabular}
    \caption{Fields and associated questions inserted into the metadata extraction prompt to retrieve specific funding information.}
    \label{tab:metadata-questions}
\end{table*}

The context of the prompt is populated dependent on the setting under evaluation i.e. baseline (the whole opportunity is included), simple (200 word chunks are inserted), or hierarchical (variable length context from the most relevant level of the document tree). Metadata fields and the associated questions to extract funding data can be found in Table~\ref{tab:metadata-questions}.

\begin{lstlisting}
system: You extract metadata from UKRI funding opportunity documents.

You will be given a document and a set of questions. Extract the answer strictly from the provided text. If the information is not explicitly stated, respond with null.

Rules:
- Only use information explicitly stated in the document.
- Monetary values should be the nearest integer pounds without currency symbols or commas.
- Percentages should be the nearest integer without the % symbol.
- Durations should be quoted using the unit they appear with in the text. Duration strings may be rewritten to make sense (e.g. "36 months", "3 years").

Respond with a JSON object only: {"<key>": "plain-text answer"} or {"<key>": null} if unknown. No other text.

user:
## CONTEXT
{context}

## QUESTION
{question}
\end{lstlisting}

\section{Council Distribution of Test Data}
\label{appendix:eval-council-distribution}

\begin{figure*}
    \centering
    \includegraphics[width=0.75\linewidth]{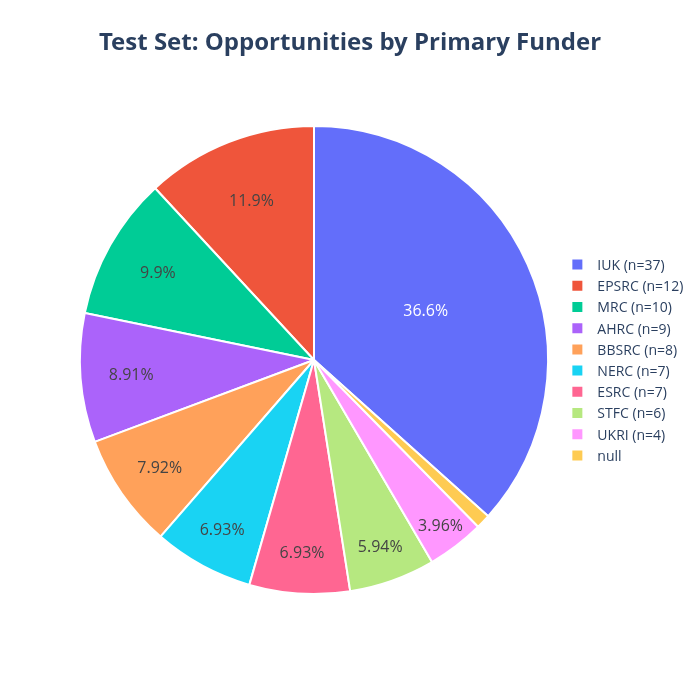}
    \caption{Breakdown of test samples by funding organisation. Samples were selected proportional to their total occurrence in the full dataset.}
    \label{fig:eval-council-distribution}
\end{figure*}

\section{Full Extraction Results}
\label{appendix:full-results}

\begin{table*}
\centering
\small
\begin{tabular}{llccccccccc}
\toprule
\multicolumn{2}{c}{Configuration} & & \multicolumn{6}{c}{Field Accuracy (\%)} & \multicolumn{2}{c}{Overall} \\
\cmidrule(lr){1-2} \cmidrule(lr){4-9} \cmidrule(lr){10-11}
Chunker & $\alpha$ & RR & Min & Max & Total & \% Funded & Min Dur & Max Dur & Acc (\%) & Unk (\%) \\
\midrule
    None & --- &  & 83.5 & 59.7 & 72.4 & 98.0 & 87.1 & 94.1 & 85.3 & 53.4 \\
    \addlinespace
    Simple & 0.00 &  & \textbf{89.4} & 59.7 & 79.3 & 94.1 & 83.2 & 86.1 & 83.4 & 58.3 \\
    Simple & 0.00 & \checkmark & 77.6 & 56.9 & 69.0 & 95.0 & 87.1 & 93.1 & 82.8 & 54.0 \\
    Simple & 0.25 &  & 82.4 & 55.6 & \textbf{82.8} & 85.1 & 85.1 & 88.1 & 80.8 & 58.7 \\
    Simple & 0.25 & \checkmark & 76.5 & 59.7 & 79.3 & 96.0 & 86.1 & 93.1 & 83.6 & 53.4 \\
    Simple & 0.50 &  & \textbf{89.4} & 61.1 & 75.9 & 78.2 & 85.1 & 88.1 & 81.0 & 63.2 \\
    Simple & 0.50 & \checkmark & 77.6 & 61.1 & 72.4 & 94.1 & 86.1 & 92.1 & 83.0 & 54.0 \\
    Simple & 0.75 &  & 84.7 & 56.9 & 79.3 & 77.2 & 84.2 & 82.2 & 78.1 & 64.4 \\
    Simple & 0.75 & \checkmark & 74.1 & 61.1 & 79.3 & 93.1 & 86.1 & 90.1 & 82.2 & 53.2 \\
    Simple & 1.00 &  & 82.4 & 55.6 & \textbf{82.8} & 78.2 & 83.2 & 78.2 & 76.9 & 65.4 \\
    Simple & 1.00 & \checkmark & 74.1 & 56.9 & 75.9 & 94.1 & 86.1 & 93.1 & 82.2 & \textbf{53.0} \\
    \addlinespace
    Hierarchical & 0.00 &  & 87.1 & 58.3 & 69.0 & 94.1 & 88.1 & 88.1 & 83.6 & 57.3 \\
    Hierarchical & 0.00 & \checkmark & \textbf{89.4} & 65.3 & 79.3 & 97.0 & 88.1 & 92.1 & 87.1 & 58.1 \\
    Hierarchical & 0.25 &  & \textbf{89.4} & \textbf{66.7} & 72.4 & 99.0 & 87.1 & 95.0 & \textbf{87.7} & 55.4 \\
    Hierarchical & 0.25 & \checkmark & 88.2 & 62.5 & 79.3 & 96.0 & \textbf{89.1} & 94.1 & 86.9 & 56.9 \\
    Hierarchical & 0.50 &  & 87.1 & 65.3 & 65.5 & 98.0 & 88.1 & 95.0 & 86.7 & 55.2 \\
    Hierarchical & 0.50 & \checkmark & 88.2 & 63.9 & 72.4 & 96.0 & 88.1 & 92.1 & 86.1 & 57.5 \\
    Hierarchical & 0.75 &  & 88.2 & 65.3 & 72.4 & \textbf{100.0} & 84.2 & \textbf{96.0} & 87.1 & 55.0 \\
    Hierarchical & 0.75 & \checkmark & 87.1 & 61.1 & 72.4 & 96.0 & 86.1 & 92.1 & 85.1 & 56.2 \\
    Hierarchical & 1.00 &  & 85.9 & 62.5 & 69.0 & 98.0 & \textbf{89.1} & 93.1 & 86.1 & 56.6 \\
    Hierarchical & 1.00 & \checkmark & 84.7 & 63.9 & 75.9 & 97.0 & \textbf{89.1} & 92.1 & 86.1 & 56.9 \\
\bottomrule
\end{tabular}
\caption{Extraction accuracy across retrieval configurations. RR = re-ranker enabled. Min/Max/Total = minimum, maximum, total award (\pounds). \% Funded = UKRI contribution percentage. Min/Max Dur = minimum, maximum funding duration. Configurations are grouped by chunking strategy; within each group, rows are ordered by $\alpha$.}
\label{tab:extraction-results}
\end{table*}

\end{document}